\documentclass[oldversion]{aamod}
\usepackage{txfonts, natbib, latexsym, graphicx, wasysym}
\bibpunct{(}{)}{;}{a}{}{,} % to follow the A&A style

\begin{document}

\title{Comment on ''The shape and composition of interstellar silicate grains''}
\author{J. P. Bradley
\and H. A. Ishii }
\institute{Institute of Geophysics and Planetary Physics, Lawrence Livermore National Laboratory, Livermore, CA 94550, USA }
\date{Accepted for publication April 4, 2008}
\titlerunning{Comment on "The shape and composition of interstellar silicate grains"}
\authorrunning{Bradley \& Ishii}
\maketitle

%\section{Introduction}
\noindent In the paper entitled ''The shape and composition of interstellar silicate grains'' ({\it A\&A}, 462, 667-676 (2007)), Min et al. explore non-spherical grain shape and composition in modeling the interstellar 10 and 20 $\mu$m extinction features.  This progression towards more physically realistic models is vitally important to enabling valid comparisons between dust observations and laboratory measurements. Min et al. proceed to compare their model results with GEMS (glass with embedded metals and sulfides) from interplanetary dust particles (IDPs) and to discuss the nature and origin of GEMS.  Specifically, they evaluate the hypothesis of ~\citet{Bradley1994} that GEMS are remnant interstellar (IS) amorphous silicates. From a comparison of the mineralogy, chemical compositions, and infrared (IR) spectral properties of GEMS with their modeling results, Min et al. conclude that ''the composition of interstellar medium (ISM) silicates is not consistent with that of GEMS'' and that ''GEMS are, in general, not unprocessed leftovers from the diffuse ISM''. These original conclusions were based, however, on erroneous GEMS data. \\
\indent In the accompanying Erratum, Min et al. provide corrections, evaluate the impact on the paper and retain their original conclusions that GEMS chemical properties are inconsistent with the bulk ISM and that most GEMS formed in the early solar system.  We respectfully disagree. With the corrections, two additional fundamental chemical properties of GEMS (Mg/(Mg+Fe) and O/Si), as well as a key optical property, the ~10 $\mu$m infrared silicate feature\footnote{Most reported GEMS infrared spectra are contaminated by neighboring crystalline and other amorphous silicates due to size mismatch between typical GEMS and instrument beam spots, but a pure GEMS infrared measurement displayed a maximum around 9.7 $\mu$m~\citep{Bradley1999} consistent with the ISM silicate feature.}, are consistent with those of the IS amorphous silicates as predicted by the Min et al. model, which itself relies on assumptions.  The additional properties further solidify the remarkable similarity between the exotic properties of GEMS and IS amorphous silicates~\citep{Bradley1994, Flynn1994, Goodman1995, Martin1995, Bradley1999}.  We agree that the match is imperfect, but given the statistical limitations of comparing less than a microgram of GEMS with the enormous mass of silicates in the ISM, this is not surprising.

\section{Interstellar GEMS versus solar system GEMS}
There is general agreement that circumstellar (CS) outflows from AGB stars are major sources of IS silicates and that at least some IS silicates are likely to retain a ÒmemoryÓ of their presolar CS origins~\citep[e.g.][]{Ebel2000}.  Whether GEMS are processed or not, the presolar IS origin (that is, extrasolar as opposed to solar system formation) of some of them has been rigorously confirmed through measurements of non-solar oxygen (O) isotope abundances~\citep{Messenger2003, Floss2006}.  In the accompanying Erratum, Min et al. acknowledge that these isotopically anomalous GEMS were indeed part of the population of IS amorphous silicates.  However, they advocate an early solar system origin for most GEMS.  Their arguments rest largely on compositional evidence.  One concerns isotopic compositions and the other concerns chemical (elemental) compositions of GEMS. 

\subsection{Isotopic composition considerations}
The isotopic argument exploits a dilemma that has long confronted meteoriticists.  Most GEMS have normal (solar) oxygen isotopic compositions suggesting that they could have formed in the solar system.  However, they could equally have formed outside of the solar system.  It is well known that only a small fraction of ISM dust can have a non-solar isotopic composition, yet a non-solar isotopic composition is currently the only way to rigorously identify presolar ISM dust.  This dilemma is eloquently articulated by the late Robert M. Walker: 
\begin{quote}
{\em ''It is (also) true that pre-existing interstellar grains might not, on the average, be very different from solar-system material.  Although there is strong evidence that some circumstellar grains with distinctive isotopic signatures have survived intact in meteorites, theoretical calculations indicate that most grains are quickly destroyed in the diffuse interstellar medium~\citep{Seab1986}.  Thus, the grains found in a protostellar gas-dust cloud may themselves consist of interstellar dust grains whose compositions have been homogenized in the interstellar medium to give compositions similar to those of solar system values.  In this connection, it is interesting to note that the isotopic compositions of galactic cosmic rays is, with some exceptions, not strikingly different from average solar system material~\citep{Wiedenbeck1984}.''}~\citep[from][]{Bradley1988}.
\end{quote}
The ISM is a giant materials recycler~\citep{Seab1986, Jones1994, Ebel2000, Westphal2004}, and it is likely no accident that presolar grains found in meteorites are a ''who's who'' of the most robust and persistent minerals (e.g. silicon carbide (SiC), corundum (Al$\sb{2}$O$\sb{3}$), diamond, graphite, etc.).   Silicate minerals, on the other hand, are more susceptible to modification and destruction by aqueous alteration, heating, shocks, and irradiation~\citep{Bradley1994, Keller1997, Westphal2004, Toppani2006, Messenger2007b}.  Fortunately, silicates are considerably more abundant than other minerals (e.g. SiC \& Al$\sb{2}$O$\sb{3}$) in the ISM which likely explains why, despite their susceptibility to modification and destruction, GEMS are to date the most abundant type of isotopically anomalous presolar grain identified in meteoritic materials~\citep{Floss2006, Nguyen2007, Messenger2007a}. \\
\indent It has been argued that even some GEMS without detectable isotope anomalies are probably presolar grains from the ISM. ~\citet{Keller2000} assert that in some cases GEMS are presolar IS silicates because of their petrographic setting, i.e. they are {\em embedded} within carbonaceous material with non-solar D/H isotopic composition. The same petrographic argument was used to argue that titanium carbide crystals in a meteorite (too small to measure isotopic composition by instruments at that time) embedded within isotopically anomalous graphite spherules are {\em ''ipso facto''} presolar grains~\citep{Bernatowicz1997}.  The advent of the NanoSIMS later proved this claim to be correct~\citep{Stadermann2005}. ~\citet{Messenger2007b} even cite {\em neighboring} isotopically anomalous ($\sp{15}$N-rich) carbonaceous material to argue that some silicates in IDPs with marginal isotope anomalies are likely presolar grains.   It follows from these arguments that the petrographic association of presolar organic material with isotopically anomalous as well as isotopically normal GEMS in IDPs strongly suggests that both types of GEMS were present together in the same presolar environment(s). \\
\indent Although a non-solar isotopic composition proves that some GEMS are IS grains, that is, they resided in the ISM, a solar isotopic composition does not prove that other GEMS did not reside in the ISM, as Min et al. acknowledge in the accompanying Erratum.   While some CS silicates retain their isotopic signatures, we may expect a greater abundance of IS silicates that have been processed in the ISM beyond ready isotopic recognition of their parentage:  The low relative abundance of GEMS exhibiting detectable isotope anomalies is consistent with the expected properties of ISM-processed grains.   Furthermore, whether a given GEMS is identified as isotopically anomalous or not depends to a large extent on the instrument used to make the measurement.  Indeed, presolar grain abundances depend on instrumental resolution as well as confidence levels used for defining a grain as anomalous.  Prior to the advent of the NanoSIMS, no presolar silicates were discovered.  With the NanoSIMS, presolar silicates were soon discovered, and ~80\% of them are GEMS in CP IDPs~\citep{Messenger2007a}. However, even with NanoSIMS the sub-nanogram masses of individual GEMS, combined with measurement sensitivity limits, mean that only relatively large isotope anomalies in large individual GEMS can be reliably identified at this time~\citep{Messenger2007b}.

\begin{table*}
%\begin{minipage}[t]{\columnwidth}
\begin{minipage}[t]{\linewidth}
\caption{Astronomical and GEMS Composition Data}
\label{table1}
\centering
\begin{tabular}{c c c c c c c}
\hline\hline
 & O/Si\footnote{Typically $<5\%$ analytical uncertainty.} & Mg/Si$\sp{a}$ & S/Si$\sp{a}$ & Fe/Si$\sp{a}$ & Isotopic composition & Reference \\
 \hline
 {\bf Solar} & & 1.01 & 0.52 & 0.9 & & ~\citet{Anders1982} \\
 {\bf Diffuse ISM} & & 1 & $<0.12$ & 0.9 & & ~\cite{Sofia2004} \\
 {\bf Average (200 GEMS)} & NR\footnote{NR $=$ not reported.} & 0.6 & 0.31 & 0.54 & & ~\citet{Keller2004} \\
 {\bf Average (42 GEMS)} & NS ($+33\%$)\footnote{NS $=$ non-soichiometric (average 33\% excess O over stoichiometry).},\footnote{Stoichiometry calculated by assigning O as MgO, SiO$\sb2$, Al$\sb2$O$\sb3$, CaO and S, Fe and Ni as sulfides and metal.} & 0.65 & 0.26 & 0.44 & & ~\citet{Ishii2008} \\
 \\
 {\bf Anomalous GEMS} & & & & & & \\
GEMS 1 & S ($-1.3\%$)\footnote{S $=$ stoichiometric (within analytical uncertainty of $\pm 10 \%$.} & 0.75 & 0.31 & 1.3 & $\delta\sp{18}$O ($\permil$) $+238\pm12$ & isotopes: ~\citet{Floss2006} \\
& S ($-9.7\%$)$\sp{e}$ & 1.1 & 0.26 & 0.44 &  $\delta\sp{18}$O ($\permil$) $+238\pm12$ & and Mg/Si: this work\\
GEMS 2 & S & 0.72 & 0.19 & 0.48 & $\delta\sp{17}$O ($\permil$) $+523\pm85$ & bulk GEMS isotopic compositions:  \\
& '' & '' & '' & '' &  $\delta\sp{18}$O ($\permil$) $-100\pm37$ & ~\citet{Keller2007a} \\ 
GEMS 3 & '' & 1.2 & 0.19 & 0.43 & $\delta\sp{18}$O ($\permil$) $+80\pm20$ & '' \\
 & '' & '' & '' & '' & $\delta\sp{18}$O ($\permil$) $+145\pm30$ & subregion of GEMS 3 above \\
 GEMS 4 & '' & 0.37 & 0.26 & 0.47 & $\delta\sp{18}$O ($\permil$) $-15\pm55$ & ~\citet{Keller2008} \\
 & '' & '' & '' & '' & $\delta\sp{17}$O ($\permil$) $+1220\pm260$ & '' \\ 
 \hline
 \end{tabular}
 \end{minipage}
 \end{table*}

\subsection{Chemical composition considerations}
The chemical argument centers on the average major element chemical compositions of GEMS, which Min et al. claim are inconsistent with ISM grain abundances derived by fitting to astronomical measurements. GEMS have too little Mg, Ca and Fe and too much S relative to ISM abundances~\citep{Keller2004}.  These average element abundances imply that GEMS were not members of the ISM grain population.  However, even isotopically anomalous GEMS known to have been members of the ISM grain population have chemical compositions that are inconsistent with ISM grain abundances (Table~\ref{table1}).  The ''remarkable similarity'' between isotopically anomalous and isotopically solar GEMS reported by~\citet{Keller2007a} and mentioned in the accompanying Erratum indicates that bulk chemical composition alone cannot be used to categorize individual GEMS according to their original formation location (in the early solar system versus a presolar (non-solar system) source).  With additional statistics over the coming years, the average bulk chemical compositions of the sub-population of isotopically anomalous GEMS can be better assessed. \\
\indent S/Si exceeds the ISM upper limit in GEMS in general and in all of the (isotopically anomalous) IS GEMS that have been measured to date in particular (Table~\ref{table1}).  (We emphasize that S is not well quantified in the ISM due to oversaturation of the absorption lines~\citep{Sofia2004}.)  Min et al. propose that GEMS may have acquired their elevated S/Si contents via gas-phase sulfidization in the collapsing solar nebula cloud or in the protoplanetary disk phase. In support of sulfidization, Min et al. report preliminary evidence by x-ray mapping that sulfides appear to be located preferentially at the outer edges of GEMS. In the accompanying Erratum, Min et al. suggest the location of sulfides in GEMS is currently an open question. We note that a preference toward exterior sulfides is unnecessary to support sulfidization, and the original supporting reference~\citep{Keller2005b} does not show that sulfides are located at the outer edges of GEMS. Furthermore, other published data report GEMS with sulfides located within their interiors~\citep[e.g.][]{Keller2005a, Dai2005, Zolensky2006} as well as GEMS with sulfides located preferentially within their interiors~\citep[e.g.][]{Bradley1994, Bradley1999, Dai2004, Westphal2004}.  While there are likely examples of GEMS with sulfides located at the edges (all other distributions having been reported), all of the published data cited above, obtained using imaging, electron diffraction, nanoprobe (x-ray) analyses, and chemical mapping, indicate that sulfides are not located preferentially at the edges of grains.  It is certainly possible that IS GEMS were indeed sulfidized in the solar nebula cloud, but experimental evidence to support this hypothesis has been sought and found absent.  In any case, whether GEMS were sulfidized in the solar nebula is tangential to the key issue of whether they are remnant presolar IS silicates. Indeed, molecular cloud formation and dissipation is a common occurrence in the ISM, and ISM grains processed, mantled and accreted in other IS molecular clouds and then reinjected into the ISM will not be readily distinguishable from those processed in our own molecular cloud.

\section{A chemical signature of ISM processing of GEMS?}
Can the chemical compositions of GEMS tell us anything about their origins and/or mechanisms of formation?  Perhaps.  Mg/Si ratios have been published for a small number of GEMS with non-solar isotopic compositions to date, and they show an intriguing trend.  They tend to be enriched relative to the mean Mg/Si ratio of 242 GEMS (Table~\ref{table1}). ~\citet{Bradley1994} proposed that the chemical properties of GEMS reflect exposure to irradiation during their prolonged lifetimes in the ISM.  Although there are undoubtedly additional erosional processes acting to modify element ratios in the ISM~\citep{Jones2000, Tielens1998, Westphal2004}, the hypothesis that irradiation is the dominant effect can be tested: Although forsterite (Mg/Si=2), enstatite (Mg/Si=1) and silicate glasses (Mg/Si unknown) have all been identified in CS outflows and/or the ISM~\citep{Molster2003, Matzel2008}, CS and IS silicates are believed to be, on average, Mg-rich~\citep[e.g.][]{Molster2003}.  Irradiation of Mg-rich silicates can cause chemical gradients and changes in the relative proportions of cations, most notably the Mg/Si ratio, via chemical fractionation effects~\citep{Bradley1994, Keller1997, Carrez2002, Westphal2004, Demyk2001, Toppani2006}.  If all GEMS were derived from the ISM, then statistically the oldest and most extensively irradiated GEMS should have the lowest Mg/Si ratios.  Conversely, the least irradiated GEMS should have the highest Mg/Si ratios, and it is these GEMS that are most likely to retain a non-solar isotopic memory of their stellar origins. In other words, IS GEMS with high Mg/Si ratios are more likely to retain detectable isotope anomalies. In the past, it has not been possible to test this hypothesis because of the small number of isotopically anomalous GEMS reported and an even smaller number with associated Mg/Si ratios reported.   Although the statistics are limited, elevated Mg/Si ratios are present in three out of four of the isotopically anomalous GEMS (Table~\ref{table1}).  One of the four has an Mg/Si ratio lower than the average, but GEMS are chemically heterogeneous on a scale of less than 100 nm~\citep{Bradley1994, Keller2008}, so Mg/Si ratios measured on single ~80 nm thick thin-sections are not necessarily representative of bulk GEMS compositions, and large sample sets will be required to yield statistically relevant conclusions.  We note that Mg/Si ratios quantified in the TEM can be accurate to $\pm$3\%. \\
\indent O/Si ratios in GEMS are another potential indicator of irradiation processing.  The least irradiated GEMS may have O/Si ratios that are approximately stoichiometric~\citep{Bradley1994, Demyk2001, Carrez2002, Toppani2006}, although light elements like O can be more difficult to quantify because of x-ray self-absorption and high O backgrounds.  O/Si ratios reported for all of the isotopically anomalous GEMS are indeed stoichiometric, although again our statistics are limited to only four GEMS (Table~\ref{table1}).  The relationship between the chemical and isotope compositions of GEMS is an exciting new avenue of investigation~\citep{Matzel2008} that requires detailed future studies of a much larger population of (isotopically anomalous) GEMS.
\section{Summary}
\noindent
The central question regarding the origin of GEMS is whether, (A) they are presolar IS amorphous silicates that survived the collapsing molecular cloud and subsequent protoplanetary disk stages of the formation of our solar system to be incorporated into IDPs~\citep{Bradley1994}, or (B) they are mostly grains formed in the solar nebula~\citep{Keller2004, Keller2007a, Keller2007b, Keller2008}.  If IS amorphous silicates survived, can we recognize them?  The answer is yes, but it is a conditional yes.  Isotopically anomalous GEMS have been identified that, during transport through the ISM and formation of the solar system, retained some portion of the isotope signatures of their formation in presolar CS environments.  These GEMS were undeniably part of the population of presolar IS silicates.  They may be analogues of the amorphous silicate grains observed in the outer disks of other young stars~\citep{vanBoekel2004}.  The simplest explanation for those observed grains is that they are IS amorphous silicates that have escaped significant heating in the (outer) accretion disks~\citep{vanBoekel2007}, and laboratory heating experiments indicate that GEMS also escaped significant heating in the solar nebula accretion disk~\cite{Brownlee2005}. \\
\indent The isotopic compositions of most GEMS are normal (solar) within the detection limits of current analytical ion microprobes, and for those not petrographically associated with presolar material, their origin may remain an open question for the foreseeable future.  However, invoking OccamÕs razor, the similarity in the properties between isotopically anomalous and normal GEMS favors presolar origin and residence in the ISM for all GEMS, especially since both are found in petrographic association with isotopically anomalous presolar organic material in IDPs.  While it is possible that some isotopically normal GEMS formed in the solar system, it would indeed be remarkable, and probably unprecedented, that a population of grains as exotic as GEMS, found in only one class of meteoritic material (CP IDPs), having a similar size distribution, mineralogy, petrography and bulk chemical composition, arose by different mechanisms at different times in the environments of different classes of stars (evolved AGB stars versus protostellar nebulae like the solar nebula). \\
\indent How exciting that we do, indeed, have samples of presolar IS amorphous silicates, one of the fundamental building blocks of solar systems.

\begin{acknowledgements}
We acknowledge funding though the NASA Cosmochemistry Program (NNH07AF99I) and Origins of the Solar System Program (NNH04AB521). We also thank M. Min and coauthors for a lively exchange that has proven invaluable in clarifying our understanding of the intriguing topics of GEMS and amorphous interstellar silicates.  This work was performed under the auspices of the U.S. Department of Energy by Lawrence Livermore National Laboratory under Contract DE-AC52-07NA27344.
\end{acknowledgements}

\bibliographystyle{aa} % style aa.bst
\bibliography{Refs} % your references in YourFile.bib

\end{document}